\documentclass[twocolumn]{webofc}
\usepackage[varg]{txfonts}   % Web of Conferences font
\usepackage[utf8]{inputenc}
\DeclareUnicodeCharacter{0233}{\'e}

\usepackage{graphicx}% Include figure files
\usepackage{dcolumn}% Align table columns on decimal point
\usepackage{bm}% bold math
\usepackage{caption}
\usepackage{subfigure}
\usepackage[dvipsnames]{xcolor}
\usepackage{cancel}

\usepackage[inline]{trackchanges} 
\addeditor{CC}
\addeditor{AZ}
\addeditor{LK}

\begin{document}
\title{Correlating the force network evolution and dynamics in slider experiments}
%
% subtitle is optionnal
%
%%%\subtitle{Do you have a subtitle?\\ If so, write it here}

\author{\firstname{Chao}
        \lastname{Cheng}\inst{1}\fnsep\thanks{\email{cc563@njit.edu}} 
\and 
        \firstname{Aghil}
        \lastname{Abed Zadeh}\inst{2}\fnsep\thanks{\email{aghil.abed.zadeh@duke.edu}}
\and
        \firstname{Lou}
        \lastname{Kondic}\inst{1}\fnsep\thanks{\email{kondic@njit.edu}} 
 %\and
 %\firstname{Third author} \lastname{Third author}\inst{3}\fnsep\thanks{\email{Mail address for last
%              author if necessary}}
%        % etc.%
}

\institute{Department of Mathematical Sciences, New Jersey Institute of Technology, Newark, NJ, 07102
\and
           Department of Physics and Department of Neurobiology, Duke University, Durham, NC, 27708
% \and
%           Last address
          }

\abstract{The experiments involving a slider moving on top of granular media consisting of photoelastic particles in two 
dimensions have uncovered elaborate dynamics that may vary from continuous motion to crackling, periodic motion, and stick-slip
type of behavior.  We establish that there is a clear correlation between the slider dynamics and the response of the force
network that spontaneously develop in the granular system.  This correlation is established by application of the persistence
homology that allows for formulation of objective measures for quantification of time-dependent force networks.  We find 
that correlation between the slider dynamics and the force network properties is particularly strong in the dynamical 
regime characterized by well-defined stick-slip type of dynamics. 
}
\maketitle
\section{Introduction}
\label{intro}

A wide range of systems exhibit intermittent dynamics as they are slowly loaded, with different dynamical regimes 
governing many industrial and natural phenomena. 
In these systems, the energy is loaded gradually with a stable configuration and then is dissipated in fast dynamics with microscopic 
and macroscopic rearrangements \cite{Sethna2001_nat}. Examples are fracture \cite{%Bonamy2008_prl
Bares2014_prl,Bares2018_natcom},
magnetization \cite{Urbach1995_prl}, and seismic activities \cite{%Bak2002_prl
Davidsen2013_prl,Bares2018_natcom} such as earthquakes, 
in which the slowly loaded energy relaxes via fast reconfiguration. This intermittent behavior has been observed in a number of granular 
experiments and simulations~\cite{Denisov2017_sr,Liu2016_prl,zadeh_pre19,murphy2019transforming}.
In analyzing such behavior, a significant progress has been reached by studying the dynamics of a slider coupled with the boundary of
a granular system. A slider can exhibit a wide variety of dynamics, including continuous flows and periodic or intermittent
stick-slip behavior~\cite{zadeh_pre19,zadeh2019crackling,ciamarra_prl10,PicaCiamarra:2009hb}.

\begin{figure}[t!]
    \centering
    \includegraphics[width=0.8\linewidth]{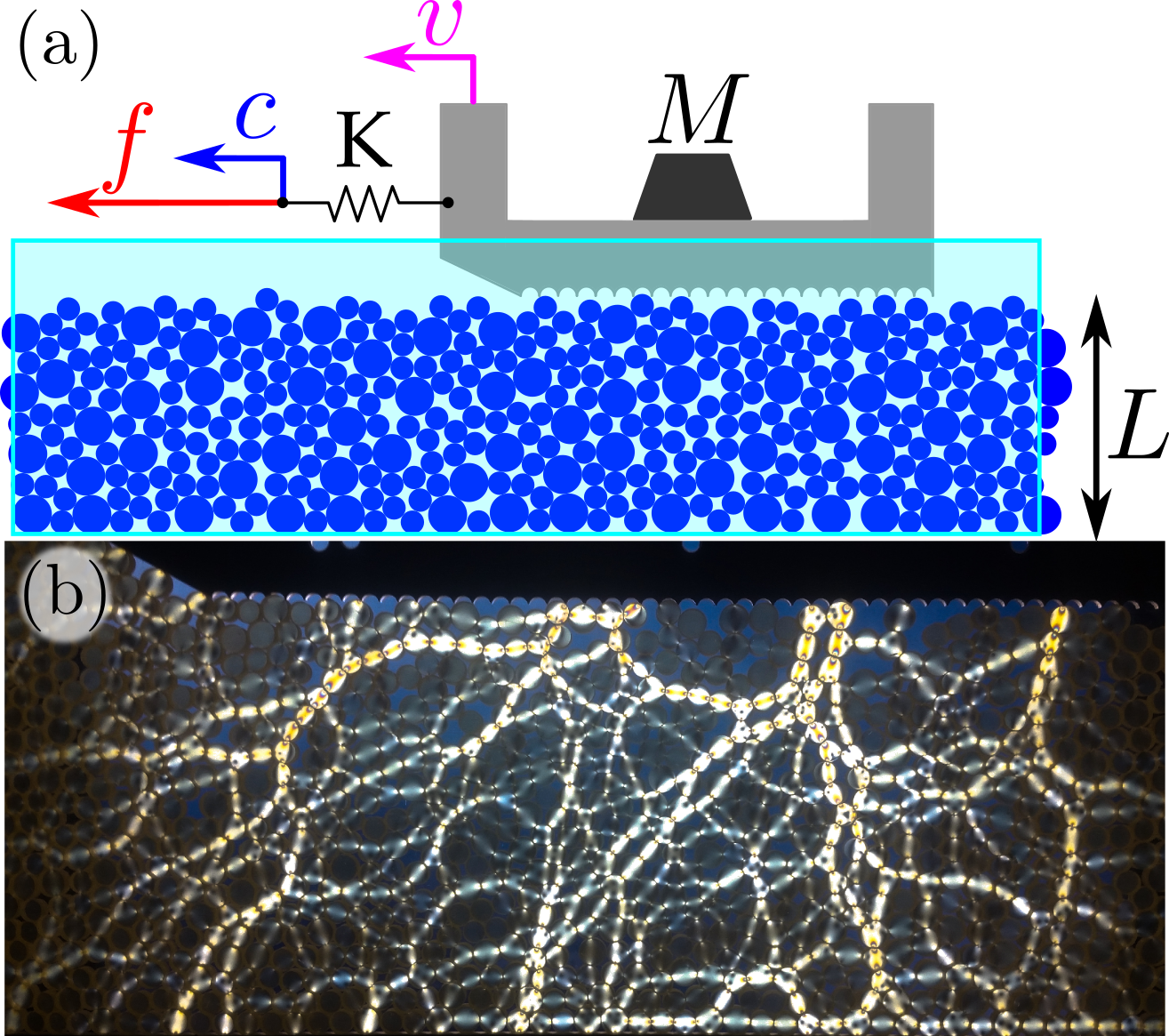}
    \caption{(a) Sketch of the experiment. A slider sits on top of a 2D granular system with photoelastic disks, and is connected to a stage by a spring of constant $k$, pulled by a constant speed $c$. A force gouge measures the force $f$; the granular medium is imaged with fast cameras.  (b) Photoelastic response during loading~\cite{zadeh2019crackling}. Reprinted with permission from~\cite{zadeh2019crackling}.}
    %\vspace{-0.3in}
    \label{fig:experiment}
\end{figure}

While a significant amount of research on exploring intermittent dynamics of granular systems has been carried out, not much is 
known about the connection between particle-scale response and the global dynamics, in particular for experimental systems.   In slider
experiments~\cite{zadeh_pre19,zadeh2019crackling}, see also Fig.~\ref{fig:experiment}, it is possible to measure particle scale response by using photoelastic 
techniques. These techniques allow for extracting dynamic information about evolving particle interactions which typically 
involve meso-scale force networks (so-called `force chains').  Analysis of such time-dependent weighted networks is not a simple task, and 
it has evolved through last decades in a variety of different directions, including force
network ensemble~\cite{tighe_sm10,sarkar_prl13}, 
statistics-based methods~\cite{peters05,makse_softmatter_2014}, and
network type of analysis~\cite{daniels_pre12, walker_pre12}.  
In the present work, we will consider application of persistent homology (PH), which allows for formulating precise and objective measures of 
static and dynamic properties of the force networks.  This approach has been used extensively in analysis of the data
obtained via discrete element simulations in the context of dry granular matter~\cite{%epl12 
ardanza_pre14, physicaD14} 
and suspensions~\cite{gameiro_prf_2020}, but its application to experimental 
data has so far been rather sparse~\cite{dijksman_2018,pre18_impact}.  We show that this method allows
to develop clear correlations between the static and dynamic properties of the force networks 
on micro- and meso-scale and the macro-scale system dynamics.

%\vspace{-0.15in}
\section{Techniques}
\label{sec-1}
\vspace{-0.1in}

\subsection{Experimental techniques}
\label{sec-2}
\vspace{-0.1in}

In our experiments, as shown in Fig.~\ref{fig:experiment}(a), a stage pulls a 2D frictional slider with toothed bottom, of a fixed length
of 25 cm and a mass of $M=$85 g . The stage, which moves at constant speed $c$, is connected to the slider by a linear spring of stiffness $K$. 
The slider rests on a vertical bed of fixed depth $L=9.5$ cm and length 1.5 m, consisting of bi-disperse photo-elastic disks with radii of 0.4 cm and 0.5 cm. A camera, recording the photo-elastic response of the medium at 120 fps, is connected to the stage. We also record the 
force $f$ experienced by the spring.

We consider three experiments characterized by different configurations of $c$ and $K$:  Exp.~1: $K= 14$ N/m, $c = 0.5$ mm/s;  
Exp.~2: $K= 70$ N/m, $c = 0.5$ mm/s, and Exp.~3: $K= 70$ N/m, $c = 1.0$ mm/s.  The total number of analyzed frames (images) is 30,000 for 
each experiment, corresponding to 250 seconds of physical time. 

\subsection{Image processing}
\label{sec-3}
\vspace{-0.1in}

\begin{figure}[t!]
    \centering
    \includegraphics[width=0.9\linewidth]{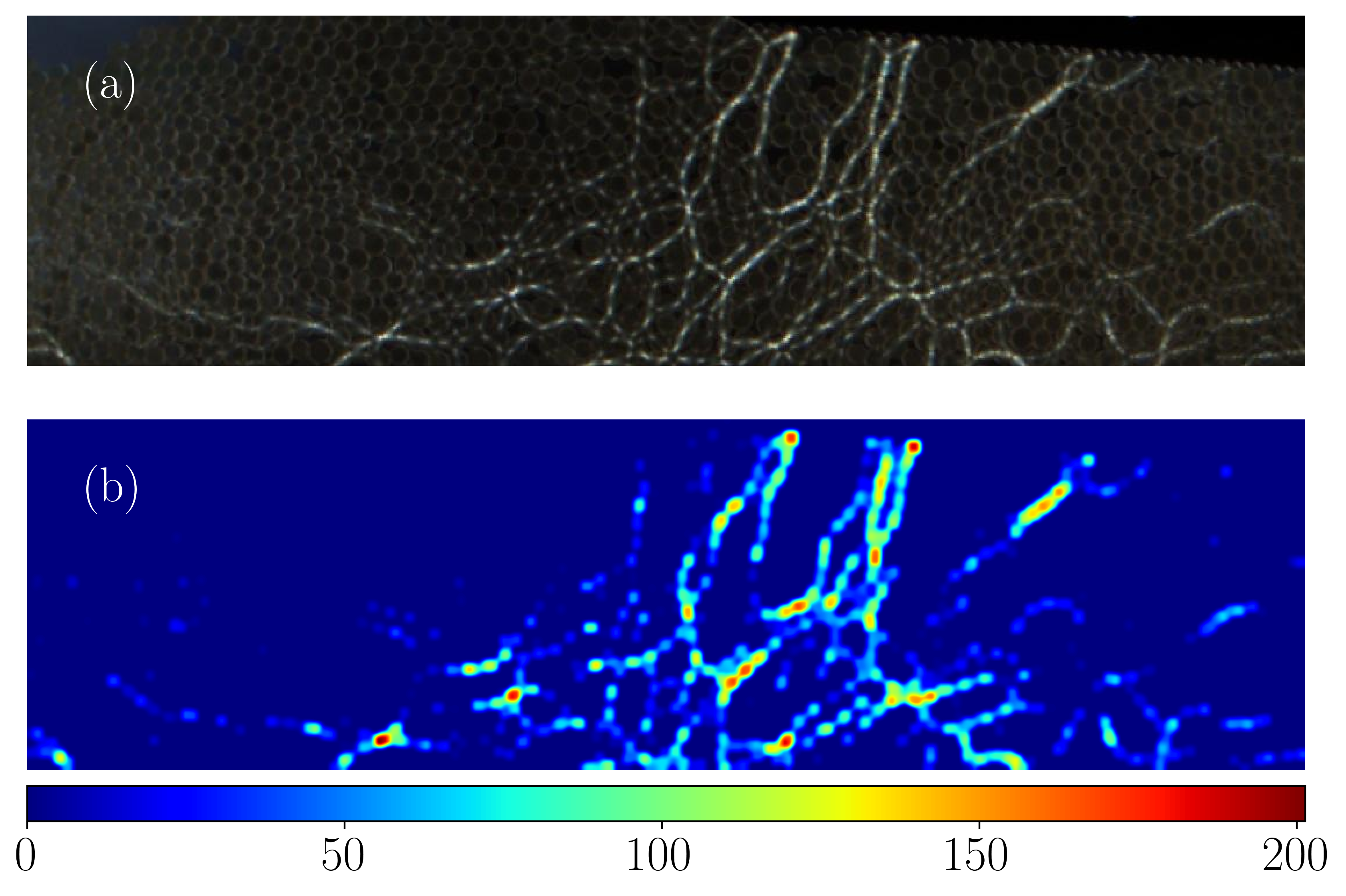}
    \label{fig:image_exp}
    \caption{An example of image processing.  (a) An original experimental image. (b) A processed image; color scheme shows pixel intensity. 
    \label{fig:image}
    }
\end{figure}

The goal of the image processing in this study is to reveal clear force signal and reduce noise effects as 
much as possible. As the fast imaging in our experiments constrains the resolution of images, we use brightness method to 
capture force information, which works better that $G^2$ method for the type of data collected, see~\cite{zadeh2019enlightening}; similar
approach was used in ~\cite{pre18_impact}. We first remove background noise from the original images by applying a filter that
removes pixels of brightness below chosen threshold value so to remove low light area and particle textures.
Multiple threshold values were investigated, giving no quantitatively difference in the results of the topological analysis that follow; 
we typically use threshold value of 90 (the maximum brightness is 255), which is appropriate 
for capturing the relevant information. After thresholding, the image brightness 
is linearly mapped to 0-255 range. MATLAB built-in functions {\em imerode} and {\em imdilate} were applied to slightly dilate the bright regions so 
to fill the gaps between neighboring particles where in force chains are connected, and then to erode away the unwanted excessive 
dilation to restore the force networks with more accuracy. Fig.~\ref{fig:image} shows an example of image processing; in our 
computations discussed in what follows, we use grey scale version of the figures such as Fig.~\ref{fig:image} for the purpose 
of computing considered topological measures.

\subsection{Topological measures}
\label{sec-4}
\vspace{-0.1in}

Persistence homology (PH) allows for formulating objective measures describing force networks in both simulations and experiments.  Analysis 
of experimental data, such as the ones considered here, presents some challenges which are discussed in some detail in~\cite{physicaD14}. 

Each experimental image can be considered as an array of pixel brightness $\theta \in [0,255]$. Since the pixels of high 
brightness correspond to 
the particles experiencing large forces, we can apply PH to the pixels to track and quantify their connectivity to 
extract (approximate) information about the actual force networks.   PH techniques essentially encode the appearance (birth)
and disappearance (death) of force networks by generating persistence diagrams (PDs) that encode the birth and death
brightness levels for components (loosely speaking, force chains), and loops (cycles).  The PDs therefore reduce the complexity 
of underlying force networks into point clouds where the coordinates are $(\theta_{\rm birth}, \theta_{\rm death})$ and each point 
represents an object that could be either connected component (chain) or a loop (cycle). The lifespan of an object is defined as
$\theta_{birth}-\theta_{death}$, measuring how long the object lasts as the threshold is varied. Total persistence (TP) of a PD 
is defined as the sum of all lifespan of the points, 
TP(PD)$=\sum_{(\theta_{\rm birth},\theta_{\rm death})\in PD}(\theta_{\rm birth}-\theta_{\rm death})$, 
which further reduce the complexity of force networks to a single number~\cite{kondic_2016}.  Note that 
TP is influenced by both how many components
there are, and by their lifespans. 

Another quantity related to PDs is the distance (or difference) between them. The distance measures essentially the cost of 
mapping points in one PD to those in another PD; in the case of different number of points, the extra ones are mapped
to the diagonal.  In particular, the \textit{degree-q Wasserstein distance} between two persistence diagrams PD and PD' is defined as 
$$
d_{Wq}(PD,PD') = \inf_{\gamma:{\rm PD} \rightarrow {\rm PD'} } \left(\sum_{p\in {\rm PD}}\|p-\gamma(p) \|_{\infty}^q \right)^{1/q},
$$
where $\gamma$ is a bijection between points from PD to PD', $\gamma:{\rm  PD} \rightarrow {\rm PD'}$. 
In the present work we use $q=2$ and carry out the calculations using the method
discussed in~\cite{TDA_exp,rTDA}.

\section{Results}
\vspace{-0.1in}
\subsection{Structural response}
\vspace{-0.1in}

%The probability density function estimate histograms indicate that smaller $c$ where the spring constant $K$ is the same, the system is more likely to be in stick-slip regime. With the same $c$, the smaller $K$ tends to have more stick-slip events.  
% \begin{figure}[ht]
%     \centering
%     \includegraphics[width=0.9\linewidth]{figs/move_mean_velocity_histogram.png}
%     \caption{Velocity distribution for different experiment}
%     \label{fig:vel_dist}
% \end{figure}
\begin{figure}[ht]
    \centering
    \includegraphics[width=1.0\linewidth]{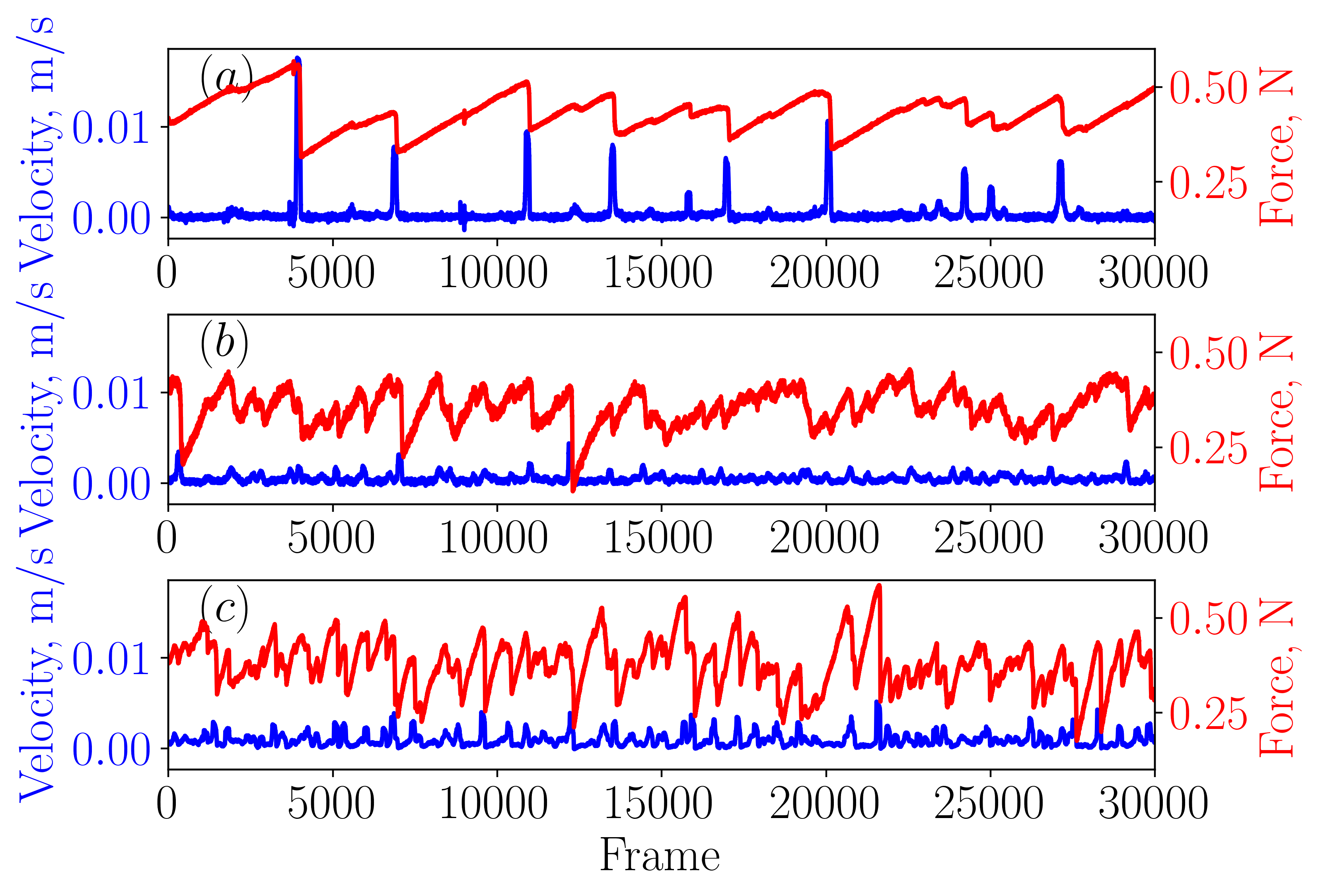}
    \caption{Sliders' velocity and spring force for the considered experiments: (a) Exp.~1, (b) Exp.~2, (c) Exp.~3.}
    \label{fig:vel_force}
\end{figure}

Figure \ref{fig:vel_force} shows the calculated velocity of the slider and the measured force,  $f$, on the spring. This figure 
illustrates clearly the slider's dynamics. We note that Exp.~1 exhibits crackling stick-slip behavior as the driving rate is small. During a stick, the spring builds up the 
stress, while the slider is almost fixed, until the spring eventually yields, leading to a sharp velocity jump 
and drop of the force. The system  behaves more similarly to a continuous flow for Exps.~2 and 3, as also
discussed in~\cite{zadeh2019crackling}. 

\begin{figure}[ht]
    \centering
    \includegraphics[width=1.0\linewidth]{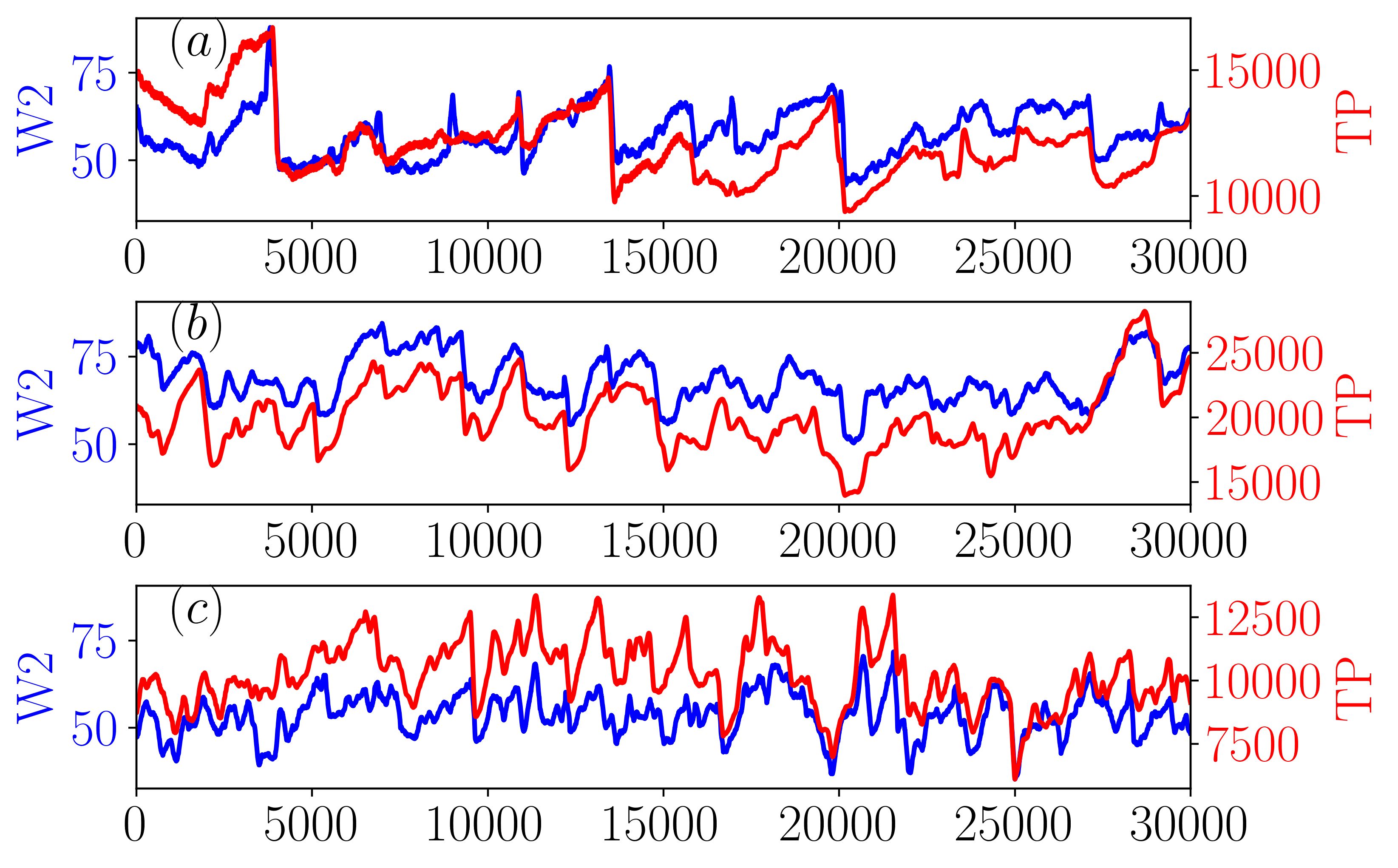}
    \caption{$W2$ distance and total persistence, corresponding to the experiments shown in 
    Fig.~\ref{fig:vel_force}.}
    \label{fig:w2_tp}
\end{figure}
\begin{figure}[t!]
    \centering
    \includegraphics[width=0.9\linewidth]{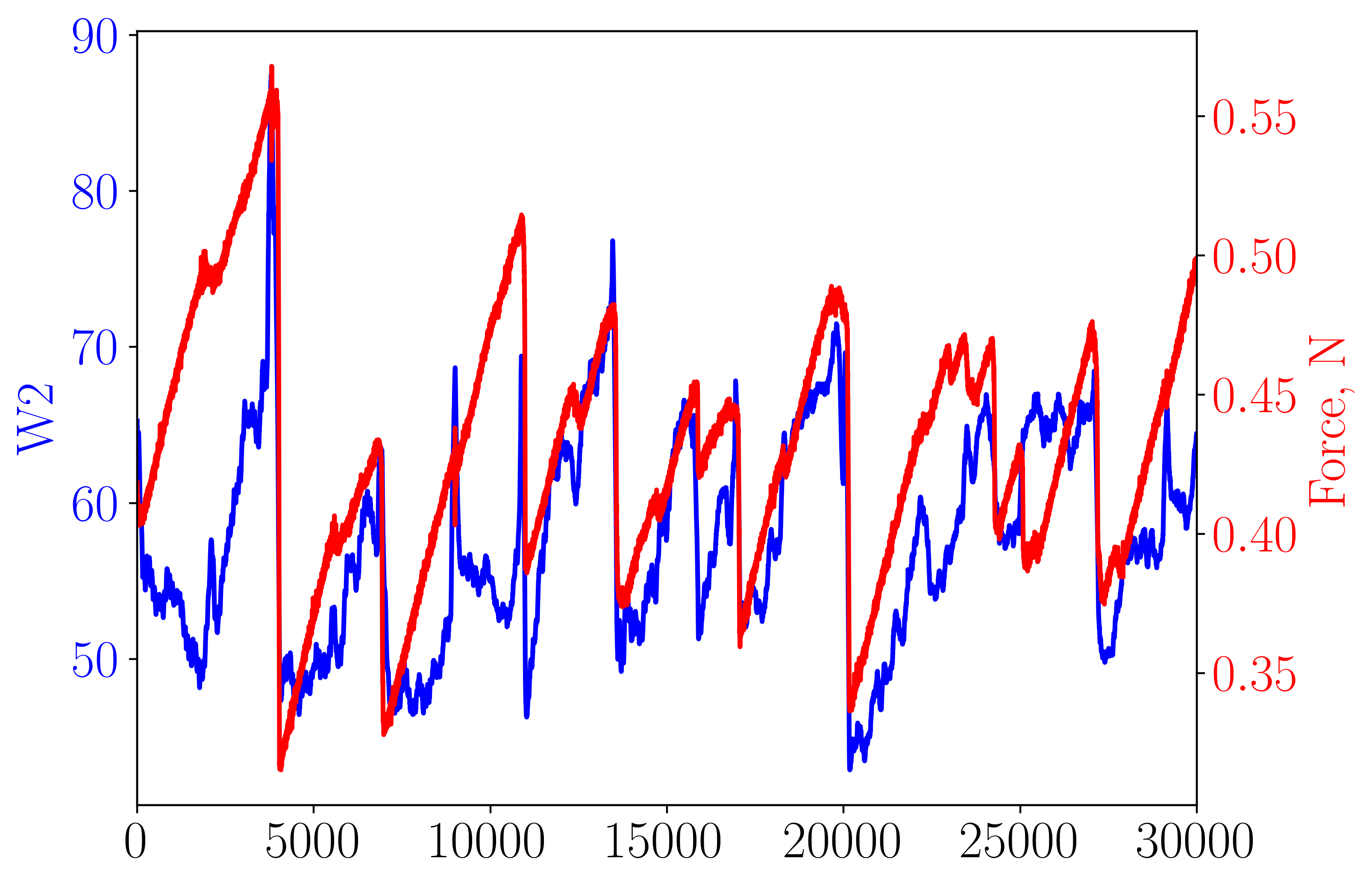}
    \caption{Force on the slider and W2 distance (data from Fig.~\ref{fig:vel_force}a) and~\ref{fig:w2_tp}a)).}
    \label{fig:force_w2}
\end{figure}

Next, we proceed with considering persistence measures.
Figure \ref{fig:w2_tp} shows the Wassertein $q=2$ distance (W2) for the considered experiments, as well as the total 
persistence (TP), for the same time interval as shown in Fig.~\ref{fig:vel_force}.    Direct comparison with Fig.~\ref{fig:vel_force} shows
good agreement between the force network measures, W2 and TP, from the one side, and the sliders' 
velocity and the spring force, on the other. Figure~\ref{fig:force_w2} illustrates in more detail the degree 
of agreement between the force and W2 for Exp.~1.  We note that this experiments shows particularly good agreement, 
suggesting stronger correlation between the force on the slider and force network properties for well defined stick-slip dynamics.
We note good agreement  between W2 and TP, 
suggesting the existence of a correlation between the `strength' of the network measured by 
TP, and its temporal evolution, measured by W2.

Having established correspondence between the slider dynamics and the force network, we proceed to discuss whether 
such correlation could be explored for predictive purposes.  To explore this question, we consider the 
cross-correlation between considered quantities.  More precisely, consider two time series 
$x_{t}$, $y_{t}$, with the data $(x_{1},x_{2},\cdots, x_{m}) $,
and $(y_{1},y_{2},\cdots,y_{m})$. The cross-covariance is defined by 
\begin{equation}
    c_{xy}(k) = \frac{1}{m}\sum_{t=1}^{m}(x_{t}-\bar{x})(y_{t-k}-\bar{y})
\end{equation}
where $\bar{x}=\sum_{i=1}^mx_i/m$, $\bar{y}=\sum_{i=1}^my_i/m$, and $k = 0, \pm 1,\pm 2,\cdots$ is the chosen lag.
When $m$ is outside the range of $y$, $y_m=0$. 
Note that for the positive lag $k$, $x_t$ is correlating with $y_{t-k}$ (at earlier time), 
which means that we may be able to use such correlation to predict the future $x$ from earlier $y$. 
Finally, we define the sample standard deviations of the series as
\begin{itemize}
    \item $s_{x} = \sqrt{c_{xx}(0)}$, where $c_{xx}(0)=Var(x)$.
    \item $s_{y} = \sqrt{c_{yy}(0)}$, where $c_{yy}(0)=Var(y)$.
\end{itemize}
The cross-correlation coefficient is given by 
\begin{equation}
    r_{xy}(k)=\frac{c_{xy}(k)}{s_{x}s_{y}}, k=0,\pm1,\pm2,\cdots
\end{equation}

\begin{figure}[ht]
    \centering
    \includegraphics[width=0.9\linewidth]{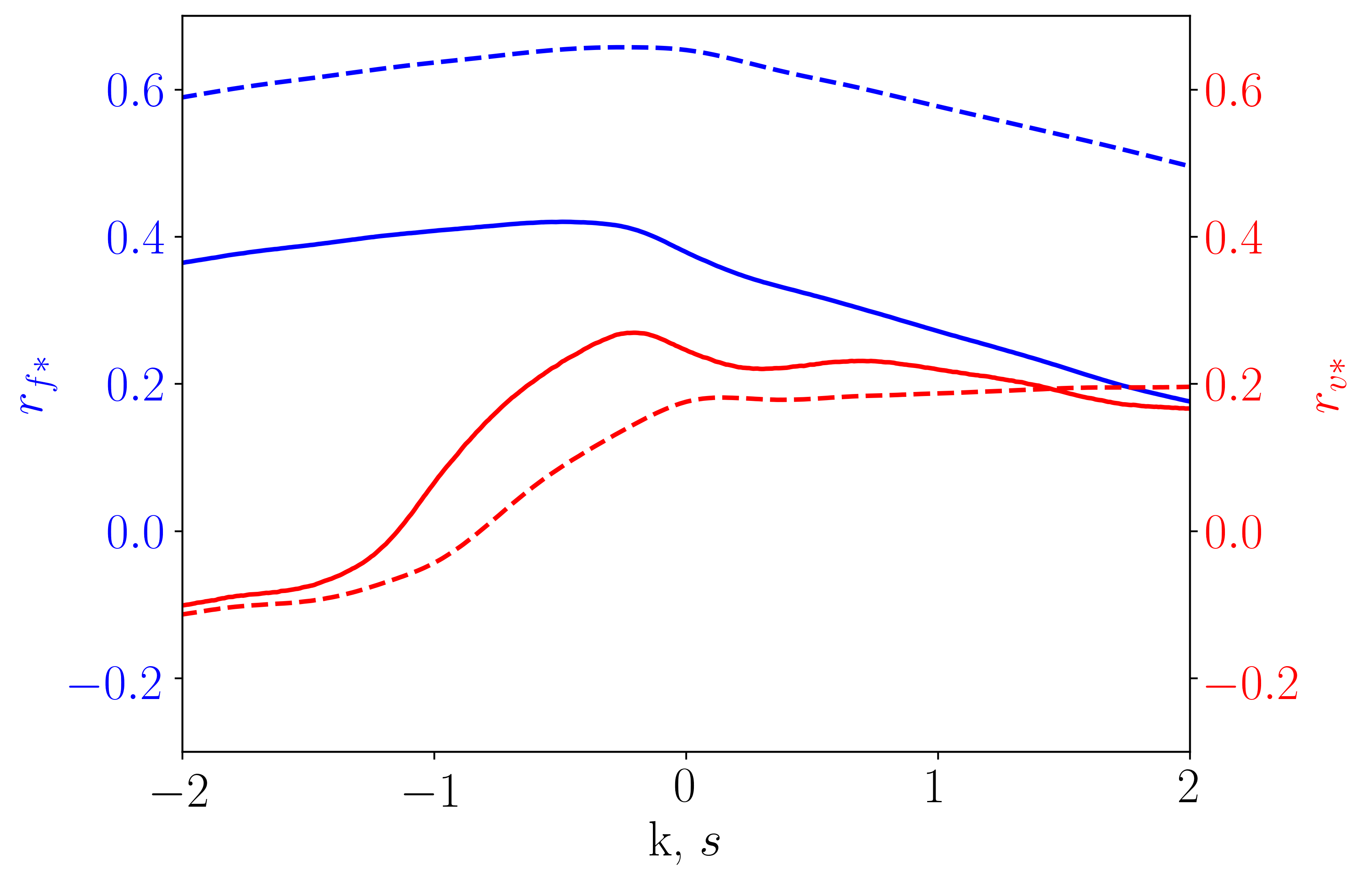}
    \caption{Cross correlation coefficient as a function of lag $k$ (in seconds). \textcolor{blue}{$r_{f*}$} and \textcolor{red}{$r_{v*}$} for Exp.~1, where $*$ stands 
    either for W2 distance (solid line),  or for TP (dashed line).}
    \label{fig:correlation}
\end{figure}

Figure \ref{fig:correlation} shows the cross-correlation coefficients for Exp.~1, where $r_{f*}$, $r_{v*}$
correspond to the cross-correlation coefficient between the $f$ or $v$ and the measure of interest
*, which can be $w$ for W2 distance or $t$ for TP. Focusing first at the results without lag, $k=0$,  
we note that the $r_{f*}$ correlations are higher than the $r_{v*}$ ones; we expect that this is due to the fact that the velocity data 
were obtained by taking a discrete derivative of the slider position data, introducing further noise 
which may blur the actual signal.
The $r_{f*}$ results show that the correlation is higher for the TP data, which is not surprising since TP is
expected to reflect the force on the slider, while W2 distance measures the temporal difference in PDs. 

Considering next the results for non-zero lags, we note a different behavior of $r_{f*}$ and $r_{v*}$ curves, with 
$r_{v*}$ curves rising from negative to positive as lag is increased. Such difference results from the fact that the structure of 
the force and velocity profiles are rather different. The velocity profile shows sharp transitions, while the force profile slowly 
builds up during the stick periods and drop dramatically at the events.  More importantly, we note that 
within a reasonable range of positive lags, the  $r_{fw}$ and $r_{ft}$ are still significant in size, suggesting a potential for predictability. It should be noted however, that since the main part of the data is in a "stick" region, the correlation will naturally weight more on these data points. A more insightful procedure would involve 
correlating the measures just before a slip event.  Such an analysis would however require more data points, 
and possibly also more detailed experimental input and therefore we leave if for the future work.   

Before closing, we note that for Exp.~2 and 3 we obtain consistent results, however the correlations are 
weaker; e.g., the max of $r_{ft}$ goes down from ~0.65 (Exp.~1), to ~0.45 (Exp.~2), and to ~0.3 (Exp.~3).
Despite the fact that all three experiments fall into the same category of stick-slip in the dynamic
phase diagram~\cite{zadeh2019crackling}, clearly (see Fig.~\ref{fig:vel_force}) the slip events are 
much stronger and better defined for Exp.~1, suggesting that in particular for such situation the
persistence measures provide insightful information.  
\vspace{-0.1in}

\section{Conclusion}
\vspace{-0.1in}

We find that the tools of persistent homology (PH) allow for correlating the dynamics of a slider 
and the photoelastic response of granular particles.  In particular stick and slip regimes of the slider dynamics are well captured by the PH measures.  These results suggest
that there is a potential for developing predictive capabilities by analyzing the response of the force network to an external 
perturbation. One open question is how precise should the information about the forces  between
the granular particles be so to allow for further development of this potential.   
We hope that our results set up a stage for this future work. 
\vspace{-0.1in}

\section*{Acknowledgements}
\vspace{-0.1in}
The authors acknowledge many insightful conversations with R. Basak,
M. Carlevaro, K. Daniels, M. Kramar, K. Mischaikow, 
J.Morris, 
L. Pugnaloni, A. Singh, J. Socolar, H. Zheng and J. Bar\'es.
CC and LK acknowledge support by the ARO grant No. W911NF1810184.
\vspace{-0.1in}
\bibliography{granulates}% Produces the bibliography via BibTeX.

\end{document}